\documentstyle[12pt,aasms4]{article}

\def\gsim{\;\rlap{\lower 2.5pt
 \hbox{$\sim$}}\raise 1.5pt\hbox{$>$}\;}
\def\lsim{\;\rlap{\lower 2.5pt
   \hbox{$\sim$}}\raise 1.5pt\hbox{$<$}\;}
\newcommand\beq{\begin{equation}}
\newcommand\eeq{\end{equation}}

\def\v{\vspace{-0.1in}}

\begin{document}

\Large
\centerline{\bf Gravitational lenses magnify up to }
\centerline{\bf one third of the most distant quasars}
\normalsize
\author{\bf J. Stuart B. Wyithe and Abraham Loeb}
\medskip
\noindent
Astronomy Dept., Harvard University, 60 Garden Street,
Cambridge, MA 02138, USA\\ 
%\centerline{submitted to {\it Nature}, February 2002}

\vskip 0.2in 
\hrule 
\vskip 0.2in

{\bf 
Exceptionally bright quasars with redshifts up to z=6.28 have recently been
discovered\cite{f1}. Quasars are thought to be powered by the accretion 
of gas onto
supermassive black holes at the centres of galaxies. Their maximum
(Eddington) luminosity is proportional to the mass of the black hole, and
so these bright quasars are inferred to have black holes with masses of
more than a few billion solar masses. The existence of such massive black
holes poses a challenge to models for the formation of structures in the
early Universe, as it requires that the black holes would grow so massive
in less than a billion years after the Big Bang. Here we show that up to a
third of known quasars with z$\sim$6 will have their observed flux 
magnified by
a factor of 10 or more through gravitational lensing by galaxies along the
line of sight. The inferred abundance of quasar host galaxies, as well as
the luminosity density provided by the quasars, are therefore substantially
overestimated.
}

The four highest redshift quasars known\cite{f1} with $z\ga5.8$ (SDSS
1044-0125 was later found\cite{d} to have $z=5.73$), were selected in the
SDSS photometric system to have magnitudes $z^*<20.2$ and colors
$i^*-z^*>2.2$. The masses of the black holes powering these quasars
are estimated to be $\ga 3\times10^9M_{\odot}$, implying\cite{hl1} that the
mass of their host galaxies is $\ga 10^{13}M_{\odot}$. Such massive hosts
lie on the steep exponential tail of the Press-Schechter\cite{PS} mass
function, so that a correction to the inferred black-hole mass severely
impacts the estimated cosmological density of galaxy halos which are
sufficiently massive to host the observed quasars.

Gravitational lensing leads to a magnification by a factor $\mu$ of the
apparent source luminosity.  Let us first consider the implications of this
magnification for the inferred properties of the massive quasar systems
within structure formation theory.  Using the Eddington luminosity to set a
lower limit on the inferred black hole mass\cite{hl1}, one finds that the
inclusion of the magnification due to lensing lowers the minimum black hole
mass by a factor of $\mu$.  This, in turn, implies that the black hole can
form in a lower mass galaxy.  Figure 1 shows the resulting enhancement
factor in the space density of galaxy halos that could host the observed
SDSS quasars at $z=6$.  The inclusion of lensing has a dramatic effect (by
orders of magnitude) on the expected abundance of such hosts at early
cosmic times.  The prominence and duty cycle of such hosts has important
implications\cite{BL1}$^,$\cite{be} with respect to the question of whether
the cosmic neutral hydrogen, a cold remnant from the big bang, was
re-ionized by star light or by quasars\cite{mhr}$^,$\cite{hl2}.

To calculate the probability of gravitational lensing in a quasar sample we
need to specify the luminosity function (number per comoving volume per
unit luminosity) that describes the quasar population. At $z\la 3$ this
function is well described by a double power-law whose shape does not
evolve with redshift\cite{p},
\begin{equation}
\phi(L,z)=\frac{\phi_\star/L_\star(z)}{[L/L_\star(z)]^{\beta_l}+[L/L_\star(z)]^{\beta_h}}.
\end{equation} 
The observed evolution of the break luminosity $L_\star$ is described by
the dependence\cite{mhr}
\begin{equation}
L_\star(z) = L_\star(0)(1+z)^{-(1+\alpha_q)}e^{\zeta z}\frac{1+e^{\xi z_\star}}{e^{\xi
z}+e^{\xi z_\star}}.
\end{equation}
We find that an intrinsic luminosity function having the parameters
$\phi_\star=624\,{\rm Gpc}^{-3}$, $\beta_l=1.64$, $\beta_h=3.43$,
$L_\star(0)=1.5\times10^{11}\,L_{\odot}$, $\alpha_q=-0.5$, $z_\star=1.45$,
$\xi=2.9$, $\zeta= 2.7$, and the inclusion of gravitational lensing
(described below) adequately describes three observables, namely the
luminosity function at $z\la 3$, and the number density of quasars with
absolute B-magnitude $M_{B}<-26$ at $z\sim4.3$ (measured by SDSS from a
catalog of 39 quasars\cite{f3} with a median redshift of $z\sim4.3$) and
$M_{B}<-27.6$ at $z\sim6.0$.  The parameter $\alpha_q$ is the slope assumed
for the typical quasar continuum $L(\nu)\propto L^{\alpha_q}$.  We are
interested in the number of quasars with luminosities higher than the
limiting magnitude $z^*_{\rm lim}=20.2$, which is $N(>L_{\rm
lim},z)=\int_{L_{\rm lim}}^{\infty}dL~\phi(L,z)$ where $L_{\rm lim}$ is the
luminosity of a quasar at redshift $z$ corresponding to an apparent
magnitude $z_{\rm lim}^*$. $L_{\rm lim}$ was determined from $z^*_{\rm
lim}$ using a luminosity distance and a $k$-correction computed from a
model quasar spectrum including the mean absorption by the intergalactic
medium\cite{mj}.

Gravitational lensing is expected to be highly probable for very luminous
quasars\cite{ov}. To illustrate this point, we consider a fictitious
gravitational lens that always produces a magnification of $\mu=4$ for the
sum of multiple images [the average value for a singular isothermal sphere
(SIS)] but $\mu=1$ otherwise. We define $\tau_{mult}$ as the probability
that a random quasar selected in the source plane will be multiply
imaged\cite{tog}, and $F_{MI}$ to be the magnification biased probability
that an observed quasar will be multiply imaged. Surveys for quasars at
$z<3$ have limiting magnitudes fainter than the break magnitude
($m_B\sim19$). While $\tau_{mult}\sim0.002$ at $z\sim2$, a survey for
quasars to a limit $L_{\rm lim}$ will find a number of lensed sources that
is larger by a bias factor of~ ${N(<L_{\rm lim}/\mu,z)}/{N(<L_{\rm
lim},z)}$.  At $z=2$, $L_{\rm lim}(z)\ll L_\star(z)$ and for $\beta_l=1.6$
the magnification bias factor is $\sim2.3$, resulting in $F_{MI}\sim0.005$.
At $z=6$, $\tau_{mult}\sim0.008$ is significantly higher\cite{t2}.
Furthermore, the limiting magnitude of the $z\ga5.8$ survey is
significantly brighter than the break magnitude and so $\beta_h=3.4$ and
the bias factor rises to $\sim28$. Under these circumstances, $F_{MI}$
rises to $\sim0.22$. These simple arguments are consistent with previous
estimates of the lensing rate at high redshift\cite{bl} and demonstrate
that lensing has a strong effect on observations of the bright SDSS quasars
at $z\ga5.8$.

To find the magnification bias more accurately, we have computed the
probability distributions $\frac{dP_{\rm sing}}{d\mu}$ and $\frac{dP_{\rm
mult}}{d\mu_{\rm tot}}$ for the magnification $\mu$ of randomly positioned
singly-imaged sources, and for the sum of magnifications $\mu_{\rm tot}$ of
randomly positioned multiply-imaged sources due to gravitational lensing by
foreground galaxies.  We assume that the lens galaxies have a constant
co-moving density [as the lensing rate for an evolving (Press-Schechter)
population of lenses differs only by $\la 10\%$ from this case\cite{bl}]
and are primarily early-type (E/S0) SIS galaxies\cite{k2} whose population
is described by a Schechter function with parameters\cite{m}
$n_\star=0.27\times 10^{-2}\,{\rm Mpc}^{-3}$ and $\alpha_s=-0.5$. We assume
the Faber-Jackson relation $(L_g/L_{g\star})=(\sigma_g/\sigma_\star)^4$
where $\sigma_g$ is the velocity dispersion of the lens galaxy, with
$\sigma_\star=220\,{\rm km\,sec}^{-1}$ and a dark matter velocity
dispersion that equals the stellar velocity dispersion\cite{k2}.  We ignore
dust extinction by the lens galaxy, which mainly arises in the much rarer
spiral galaxy lenses.  Potential lens galaxies must not be detectable in
the survey data used to select the objects.  Galaxies having $i^*<22.2$
(around 30\% of the potential lens population) are not considered part of
the lens population. To compute $i^*$ for a galaxy having velocity
dispersion $\sigma$ at redshift $z$, we use $L_{g\star}$ from the 2dF
early-type galaxy luminosity function\cite{m}, the Faber-Jackson relation,
color transformations with a $k$-correction\cite{bln}$^,$\cite{f}, and the
evolution of the rest-frame B-band mass-to-light ratio\cite{kt}.

Our lens model includes microlensing by the population of galactic stars
which is modeled as a de Vaucouleurs profile of point masses embedded in
the overall SIS mass distribution.  The surface mass density in stars for
galaxies at $z=0$ is normalized so that the total cosmological density
parameter of stars\cite{Fuk} equals $0.005$. At $z>0$ the total mass
density in stars is assumed to be proportional to the cumulative
star-formation history\cite{h}$^,$\cite{n}.  The parameters of the
de~Vaucouleurs profiles are taken from a study of the fundamental
plane\cite{dd}, the microlens mass is chosen as 0.1$M_{\odot}$ and the
source size as $10^{15}$cm (corresponding to ten Schwarzschild radii of a
$3\times 10^8M_\odot$ black hole).  The probabilities $\frac{dP_{\rm
sing}}{d\mu}$ and $\frac{dP_{\rm mult}}{d\mu_{\rm tot}}$ closely resemble
the standard form for the \emph{SIS} and were computed\cite{wt} by
combining numerical magnification maps with the distribution of
microlensing optical depth and shear along lensed lines of sight, and the
distribution $\left[\tau_{\rm mult}\frac{dP_{\rm mult}}{d\mu} +(1-\tau_{\rm
mult})\frac{dP_{\rm sing}}{d\mu}\right]$ was normalized to have unit mean.

The fraction of sources which are multiply imaged due to gravitational
lensing is
\begin{equation}
F_{\rm MI}(z)=\frac{\int_{0}^{\infty}d\mu'\tau_{\rm mult}\frac{dP_{\rm
mult}}{d\mu'}N(>\frac{L_{\rm
lim}}{\mu'},z)}{\int_{0}^{\infty}d\mu'\left[\tau_{\rm mult}\frac{dP_{\rm
mult}}{d\mu'}+ (1-\tau_{\rm mult})\frac{dP_{\rm
sing}}{d\mu'}\right]N(>\frac{L_{\rm lim}}{\mu'},z)},
\end{equation}
where $\tau_{\rm mult}=0.0059$.  We find a value of $F_{\rm MI}\sim0.30$
for the color-selected, flux-limited $z\ga5.8$ sample.  This value is
higher by two orders of magnitude than the lens fraction at low redshifts
and demonstrates that lensing must already be considered in a sample with
only 4 objects. For comparison, our model (including the entire lens
population of lens galaxies) predicts $F_{MI}\sim0.01$ at $z\sim2$ for
$m_B<20$. These calculations do not include selection effects for flux
ratios (though the large magnification bias will favor flux ratios near 1)
or image separations which may serve to lower $F_{MI}$.

A sub arc-second resolution K-band image has been obtained for the $z=5.80$
quasar\cite{f1}, and it was found to be an unresolved point source. To our
knowledge this is the only $z\ga5.8$ quasar for which sub arc-second
resolution optical imaging is currently available. However a program to
image these quasars with the {\it Hubble Space Telescope} which will
determine the multiple-image fraction, is expected to begin within the
coming year (X. Fan private communication).  Note though that single image
quasars might still be magnified by a factor of $\sim2$. Recent Chandra
observations\cite{sw} of the $z=6.28$ quasar show photons
detected on the edge of the extraction (1.2'') circle, offering a hint that
this quasar may be lensed.

We have computed the distribution of magnifications observed for a
sample of quasars brighter than $L_{\rm lim}$ at redshift $z$,
\begin{equation}
\frac{dP}{d\mu_{\rm obs}} = \frac{\left[\tau_{\rm mult}\frac{dP_{\rm
mult}}{d\mu}\vert_{\mu=\mu_{\rm obs}} + (1-\tau_{\rm mult})\frac{dP_{\rm
sing}}{d\mu}\vert_{\mu=\mu_{\rm obs}}\right]N(>\frac{L_{\rm lim}}{\mu_{\rm
obs}})}{\int_0^{\infty}d\mu'\left[\tau_{\rm mult} \frac{dP_{\rm
mult}}{d\mu}\vert_{\mu=\mu'} + (1-\tau_{\rm mult})\frac{dP_{\rm
sing}}{d\mu}\vert_{\mu=\mu'}\right]N(>\frac{L_{\rm lim}}{\mu'})}.
\end{equation}
In Figure~2 we show the probability that the magnification of a quasar is
higher than $\mu_{\rm obs}$, assuming that the quasar belongs to a sample
at a redshift $z\sim6$ with the SDSS magnitude limit of $z^*<20.2$.  The
plotted distribution is highly skewed; the median magnification is
$med(\mu_{obs})\sim1.2$ while the mean magnification is as high as
$\langle\mu\rangle=24$. Thus, one or more of the $z\ga5.8$ quasars are
likely to be highly magnified while most should be magnified at a low
level.

The large values of the magnifications and the highly skewed shape of the 
distributions in Figure~2 suggest that lensing must alter the observed
luminosity function. Indeed, we find that the space density of quasars 
with $M_{B}<-27.6$ is increased by 40\%, and that the slope is decreased 
by 0.15. A quasar magnified by $\mu$ is detectable to luminosities as low 
as $L_{lim}/\mu$ and has its luminosity, $L$, overestimated
by $\mu$. The factor $R_{LD}$ by which the luminosity density of quasars 
brighter than $L_{lim}$ is overestimated due to lensing is therefore
\begin{equation}
R_{LD} = \frac{\int_0^{\infty}d\mu'\left[\tau_{\rm mult}\frac{dP_{\rm
mult}}{d\mu'}+ (1-\tau_{\rm mult})\frac{dP_{\rm
sing}}{d\mu'}\right]\int_{L_{lim}/\mu'}^{\infty}dL'\mu'L'\phi(L')}{\int_{L_{lim}}^{\infty}dL'L'\phi(L')}
\end{equation}
We find $R_{LD}\sim2$, implying that naive computation of the quasar
luminosity density from the $z\ga5.8$ sample might significantly
overestimate its true value.  Magnification bias also affects quasars that
are not multiply imaged. We therefore predict an enhanced angular
correlation on the sky between $z\sim 6$ quasars and foreground galaxies.

The high source redshifts imply image separations that are slightly larger
than usual\cite{bl}, $\sim 1$--2$^{\prime\prime}$.  Furthermore, the lenses
are likely to be found at higher redshift\cite{bl} due in part to the
higher source redshifts but also because bright (low-$z$) lenses are
excluded.  Lensing of high-redshift quasars allows measurement of the
masses of the lens galaxies\cite{kt} at redshifts higher than currently
possible. Furthermore, quasar microlensing should be common over a ten year
baseline\cite{wt}. This offers the exciting possibility of measuring the
source size, and hence black-hole mass, which in turn yields the ratio
between quasar luminosity and its Eddington value, a quantity that will be
useful in constraining models of structure formation\cite{hl1}.

The recently published spectra of the highest redshift quasar
\cite{be}$^,$\cite{px} limit the flux in the Gunn-Peterson\cite{GP} trough 
to less than $3\times10^{-19}\,{\rm erg\,sec^{-1}\,}$\AA$^{-1}$. Given the
importance of this first observation of the re-ionization epoch and of
future complimentary observations, we are motivated to ask whether or not
lens galaxy light will contaminate deep observations of the Gunn-Peterson
trough in the quasar spectrum, even though such a galaxy is not detected in
the initial imaging survey. We have computed the flux of lens galaxies at
redshift $z$ with velocity dispersion $\sigma$, and convolved the results
with the joint probability distribution for the lens galaxy redshift and
velocity dispersion. We find that $\sim 40\%$ of multiple image lens
galaxies ($i^*<22.2$) will contribute flux in the Gunn-Peterson trough
above a level of $3\times10^{-19}\,{\rm erg\,sec^{-1}}\,$\AA$^{-1}$.  For
some quasars the contamination of the Gunn-Peterson trough by flux from
lens galaxies may limit the ability of deep spectroscopic observations to
probe the evolution of the neutral hydrogen fraction during the epoch of
reionization.

\vskip 0.45in

\noindent {\it Correspondence and requests for materials to Abraham Loeb.}

\vskip 0.45in
\hrule
\vskip 0.2in
\noindent
{ACKNOWLEDGEMENTS.} The authors would like to thank Ed Turner and Josh Winn
and Rennan Barkana for helpful comment and discussion. This work was 
supported in part by grants from the NSF and NASA. J.S.B.W. is supported 
by a Hubble Fellowship.

\vskip 1in

\begin{figure*}[hptb]
\epsscale{.5}
\plotone{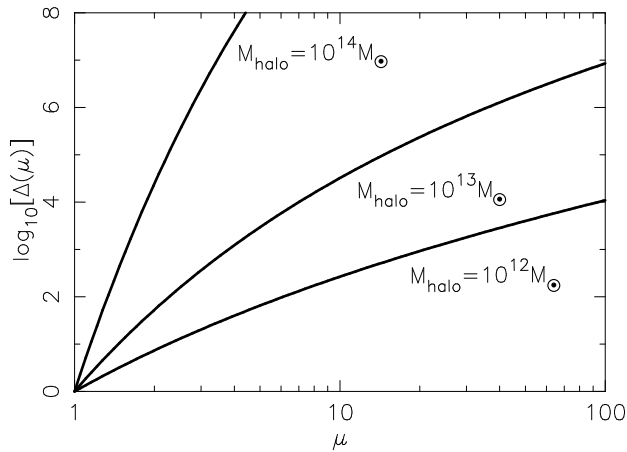}
\caption{Enhancement in the co-moving space density of host galaxies for
quasars at $z=6$ as a function of the magnification due to lensing, $\mu$.
Ignoring lensing, the Eddington limit implies\cite{hl1} a minimum black
hole mass of $\sim 3\times 10^9M_\odot$ for the SDSS quasars at $z\sim 6$.
The inclusion of lensing reduces the implied (minimum) black hole mass by a
factor of $\mu$. Given some fixed efficiency for assembling gas into a
central black hole within a galaxy, the implied mass of the host galaxy is
also lowered by $\mu$ after lensing is included. Plotted is
$\Delta(\mu)=\frac{dn}{d(log[M/\mu])}/\frac{dn}{d(logM)}$, where
$\frac{dn}{d(logM)}$ is the Press-Schechter\cite{PS} comoving density of
galaxy halos with mass $M$ per logarithmic interval in $M$. We show curves
for three possible halo masses of quasar hosts,
$M_{halo}=10^{12}M_{\odot}$, $10^{13}M_{\odot}$, and $10^{14}M_{\odot}$. As
seen, the enhancement in the implied abundance of quasar hosts increases
with increasing halo mass.  The above three choices for halo masses are all
{\it conservatively} lower than inferred for the same black hole mass in
the local universe\cite{fe}.  Throughout this {\it Letter} we assume the
standard cosmological parameters $\Omega_{m}=0.35$,
$\Omega_{\Lambda}=0.65$, $\Omega_b=0.05$, $H_0=65\,{\rm
km\,sec^{-1}\,Mpc}$, $n=1$ and $\sigma_8=0.8$.}
\end{figure*}

\begin{figure*}[hptb]
\epsscale{.4}
\plotone{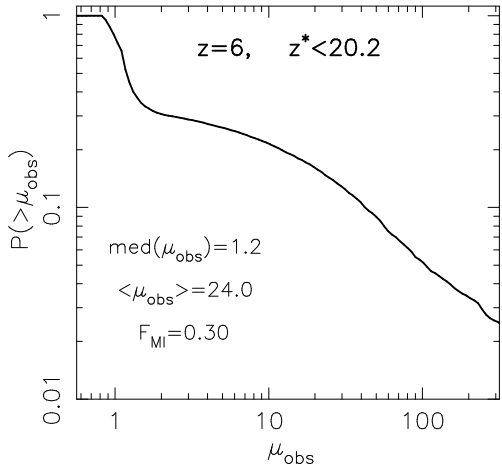}
\caption{The probability of observing a magnification larger than $\mu_{\rm
obs}$ for a quasar at a redshift $z=6$ in a sample with a magnitude limit
$z^*<20.2$. The distribution is highly skewed, having a median of
$med(\mu_{obs})=1.2$ and a mean of $\langle\mu_{obs}\rangle=24.0$. The
multiple image fraction is $F_{MI}=0.30$. We have also computed the {\it
a-posteriori} values of $F_{\rm MI}$ and $\langle\mu\rangle$ for specific
quasars.  For SDSS~0836-0054 ($z=5.82$), SDSS~1306-0356 ($z=5.99$) and
SDSS~1030-0524 ($z=6.28$) we find $F_{\rm MI}=0.40$, 0.32 and 0.31, and
$\langle\mu\rangle=50$, 25 and 23, respectively.  While Fan et al.\cite{f1}
find that $\beta_h=3.43$ is consistent with the luminosities of the
$z\ga5.8$ quasars, it is possible that the luminosity function at high
redshift is not as steep as $\beta_h=3.43$.  In this case, the
magnification bias will not be as large and the mean magnification and
fraction of quasars that are multiply imaged will be lower. We have
re-computed the lens statistics assuming that the bright end slope is
significantly flatter at high redshift.  Assuming that $\beta_l=1.64$ at
all redshifts and $\beta_h=3.43$ for $z<3$ but $\beta_h=2.58$ for $z>3$
(the value found\cite{f3} for quasars at $z\sim4.3$) we inferred similar
parameters to describe the observed luminosity function as before
[$L_{\star}(0)=1.5\times10^{11}L_\odot$, $z_\star=1.6$, $\xi=3.3$, $\zeta=
2.65$]. Remarkably, the multiple image fraction is still nearly 0.1 in this
case. Additional uncertainty in the calculation arises in the choice of
lens model. The value $\tau_{mult}$ is proportional to
$n_\star\sigma_\star^4$. The dependence of $F_{MI}$ on $\tau_{mult}$ is
complex [see equation~(3)], however large magnification biases result in a
relation that is less sensitive than linear. For example, reducing the
value of $\tau_{mult}$ by a factor of 1.5 results in $F_{MI}=0.22$ if
$\beta_h=3.43$ and $F_{MI}=0.043$ if $\beta_h=2.58$.}
\end{figure*}

\end{document}